# Performance Characteristics of the Battery-Operated Si PIN Diode Detector with Integrated Preamplifier and Data Acquisition Module for Fusion Particle Detection

Allan X. Chen [a]*, Benjamin F. Sigal [a], Qiong Wang [a], John Martinis [a], Naomi Mitchell [a], Yuxing Wang [a], Alfred Y. Wong [b], Zhifei Li [b], Alexander Gunn [b], Matthew Salazar [b], Nawar Abdalla [b], Benjamin Wrixon [b], Chia-Yi Chen [c], Nai-Wei Liu [c], KaiJian Xiao [c], Chih-Jui Hsieh [c], Ming-Cheng Jheng [c]

Alpha Ring International, Limited.

[a] 1631 W. 135th St. Gardena, CA 90249, U.S.A.

[b] 5 Harris Ct. Suite B, Monterey, CA 93940, U.S.A.

[c] 9F., No.11, Ln.35, Jihu Rd. Neihu Dist. Taipei City 114066, Taiwan

**Abstract**

We present the performance and application of a commercial off-the shelf Si PIN diode (Hamamatsu S14605) as a charged particle detector in a compact ion beam system (IBS) capable of generating D-D and p-B fusion charged particles. This detector is inexpensive, widely available, and operates in photoconductive mode under a reverse bias voltage of 12 V, supplied by an A23 battery. A charge-sensitive preamplifier (CSP) is powered by two 3 V lithium batteries (A123), providing ±3 V rail voltages. Both the detector and preamplifier circuits are integrated onto the same 4-layer PCB and housed on the vacuum side of the IBS, facing the fusion target. The system employs a CF-2.75 flanged DB-9 connector feedthrough to supply the signal, bias voltage, and rail voltages. To mitigate the high sensitivity of the detector to optical light, a thin aluminum foil assembly is used to block optical emissions from the ion beam and target. Charged particles generate step responses on the preamplifier output, with pulse rise times on the order of 0.2 to 0.3 µs. These signals are recorded using a custom-built data acquisition unit, which features an optical fiber data link to ensure electrical isolation of the detector electronics. Subsequent digital signal processing is employed to optimally shape the pulses using a CR-RC$^n$ filter to produce Gaussian-shaped signals, enabling accurate extraction of energy information. Performance results show that the signal-to-noise ratios (S/N) for D-D fusion charged particles—protons, tritons, and helions—are approximately 30, 10, and 5, respectively, with a shaping time constant of 4 µs.

## Introduction

Si PIN diode and other electronic-based charged particle detectors have gained prominence in recent years for fusion particle detection [1][2]. Unlike solid-state nuclear track detectors (SSNTDs) such as CR-39, PIN detectors provide a real-time, high-fidelity method for particle identification and energy measurement. With an appropriate preamplifier and pulse-shaping amplifier, high energy resolution can be achieved, with full-width-half-maximum (FWHM) energy resolutions of 10-20 keV for Am-241 alpha particles reported in the literature [3]. In fusion environments, proper shielding against optical light and protection of electronics from electromagnetic interference (EMI) are essential for reliable operation. To address these challenges, Alpha Ring International developed a compact fusion demonstration system that uses an H+ or D+ ion beam to produce D-D and p-B11 fusion through the beam-target method [1]. The system allows for precise control of beam parameters (e.g., energy, current, pulse length), creating a highly controlled environment for testing and calibrating detectors with real fusion particles. This study focuses on the Hamamatsu S14605 PIN diode [4], selected for its low cost and ready availability. As an OEM module, the diode requires integration with a custom-designed preamplifier PCB for optimal performance (Fig. 1). To reduce costs and maintain a compact form factor, we developed a custom data acquisition system integrated with the detector and preamplifier. The result is a self-contained unit that can be controlled with a USB-UART port, providing a versatile and cost-effective solution for fusion particle detection.

## Detector System Design

The S14605 Si PIN detector has an active area of 9mm x 9mm with a nominal depletion depth of 500μm. A reverse bias is crucial to obtain good detector rise time and low baseline noise. Additionally, the pulse rise time is comparatively faster with a detector bias due to the presence of a sweeping electric field to carry away the charges quickly. The detector can operate with a bias voltage as high as +150V, however, we have found adequate performance operating with a bias voltage of +12V, which can be easily supplied with an A23 battery [5].

The PIN detector readout circuit (Fig. 2) utilizes a JFET input op-amp (ADA4625-2) as the transimpedance amplifier (TIA) and a subsequent gain=2 non-inverting buffer. This op-amp exhibits very low input voltage and current noise, which is crucial for the charge-sensitive nature of the detector. Based on the specification sheet [6], the 1/f voltage noise knee frequency is ~200 Hz and noise density ~3.3 nV/Hz$^{1/2}$. The low knee frequency is ideal for detection of the fast rise-time pulses as the noise occupancy at the higher frequency is minimized. Figure 2 shows the circuit diagram of the readout electronics. Note that the rail voltages are set to ±2.5 V, which are supplied via. positive and negative voltage linear regulators. The input of the regulators is powered from 3-V Li batteries (A123) [7]. These batteries offer very high charge capacity for their size, so the data acquisition can be operated for a long time before replacement of the batteries. Due to the high impedance of the PIN diode in reverse bias, the A123 12-V biasing battery does not consume much charge and typically last much longer than the 3 V batteries powering the electronics.

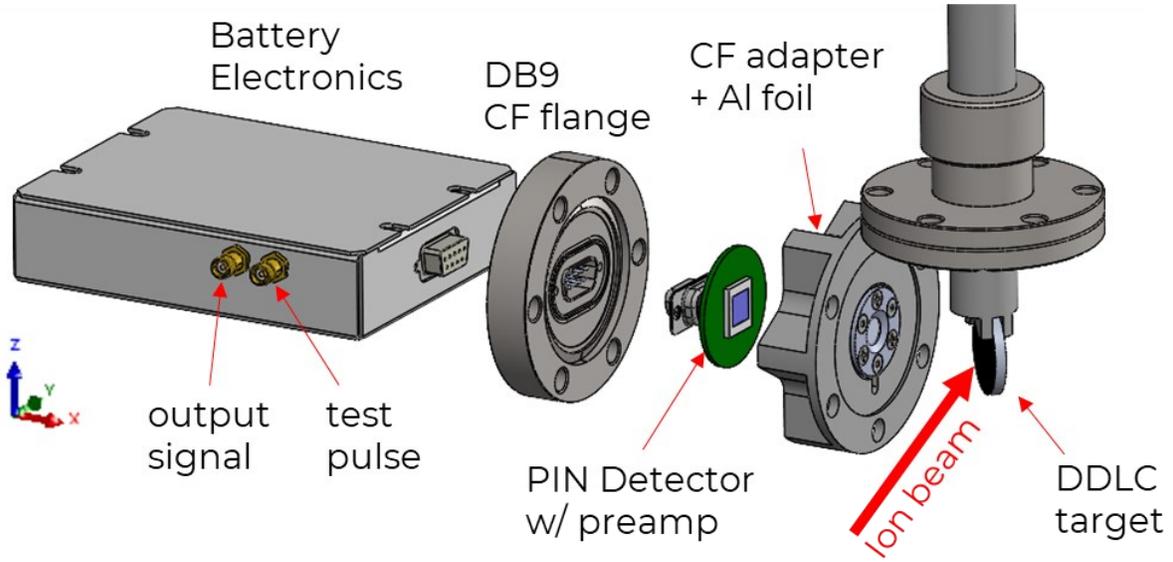

*Figure 1: Detector system hardware stack showing how the Si-PIN detector is mounted in vacuum to observe fusion charged particles from a beam-target experiment.*

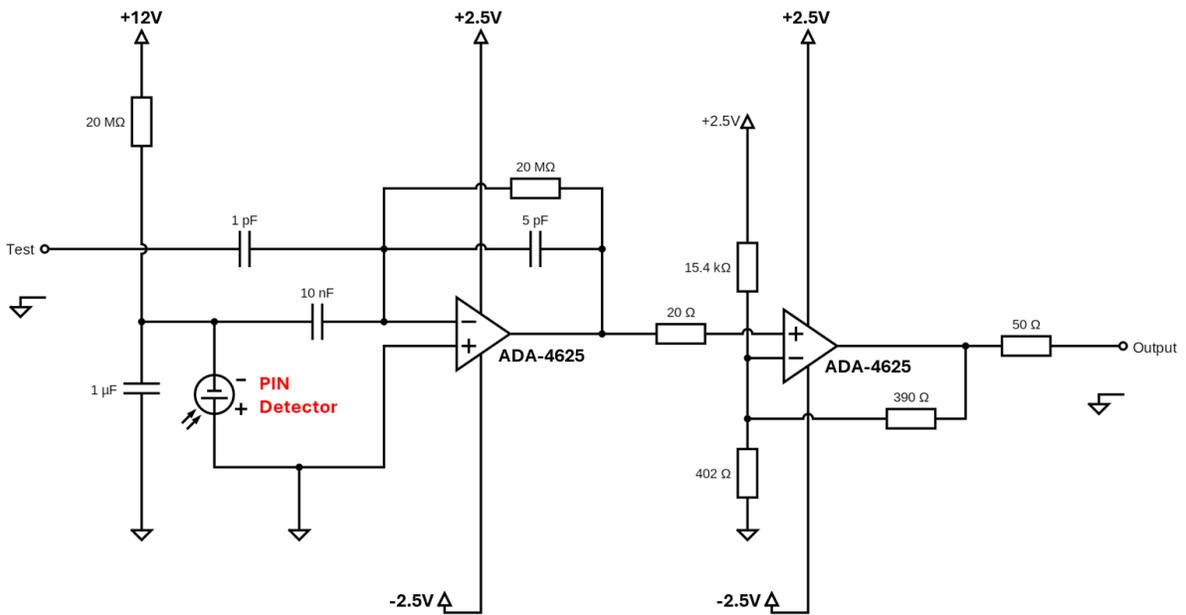

*Figure 2: Electronics schematic of the Si-PIN detector and preamplifier.*

**Signal Analysis Method and Results**

The raw preamplifier signals exhibit a fast rise time of approximately 0.3–0.6 μs, followed by a long decay time determined by the RC constant of the feedback capacitor and resistor on the first operational amplifier, which is approximately 10 ms. To process these signals, discrete-time filtering methods based on the CR-RC$^4$ configuration, as described in [1], have been employed. A typical pulse-shaping time of 0.5 to 2 μs is used, balancing signal integrity with the minimization of pile-up events. This filtering approach is effective for observing fusion products at lower beam voltages, where beam-induced noise is minimal. However, at higher beam voltages, the induced noise becomes significant, often comparable in magnitude to the charged particle signal. For example, Figure 3 demonstrates the detection of the p-B$^{11}$ alpha-particle signal from a system operating at 90 keV, 5 kHz, and a 5% duty cycle. In this scenario, the fast-rising alpha-particle signal is embedded within a periodic noise signal induced by the beam. While the alpha-particles can still be distinguished from noise based on their rise-time constant, the resulting spectrum exhibits a pronounced noise peak in the low-energy region (Figure 4).

To address this, a CR$^2$-RC$^4$ discrete-time filtering method was applied using the same pulse-shaping time constant. This approach yielded a significantly cleaner filtered bipolar signal. The resulting energy spectrum displayed a drastically reduced noise peak (Figure 4), enabling the detection of the low-energy p-B$^{10}$ alpha peak. The $^{10}$B(p,α)$^7$Be reaction has a Q-value of 1.1 MeV, imparting an alpha-particle kinetic energy of 0.7 MeV. Based on SRIM stopping power calculations for a 0.8 μm foil, the observed peak shift is consistent with the calculated energy spectrum. To our knowledge, this is the first time the proton-boron fusion energy spectrum has been obtained using a natural-abundance boride target, clearly showing the p-B$^{10}$ reaction alpha-particle. In contrast, previous measurements of the alpha-particle energy spectrum, such as those reported in [8], were conducted at an H$^+$ energy of 675 keV, which would have masked the p-B$^{10}$ alpha-particle peak due to scattered beam protons.

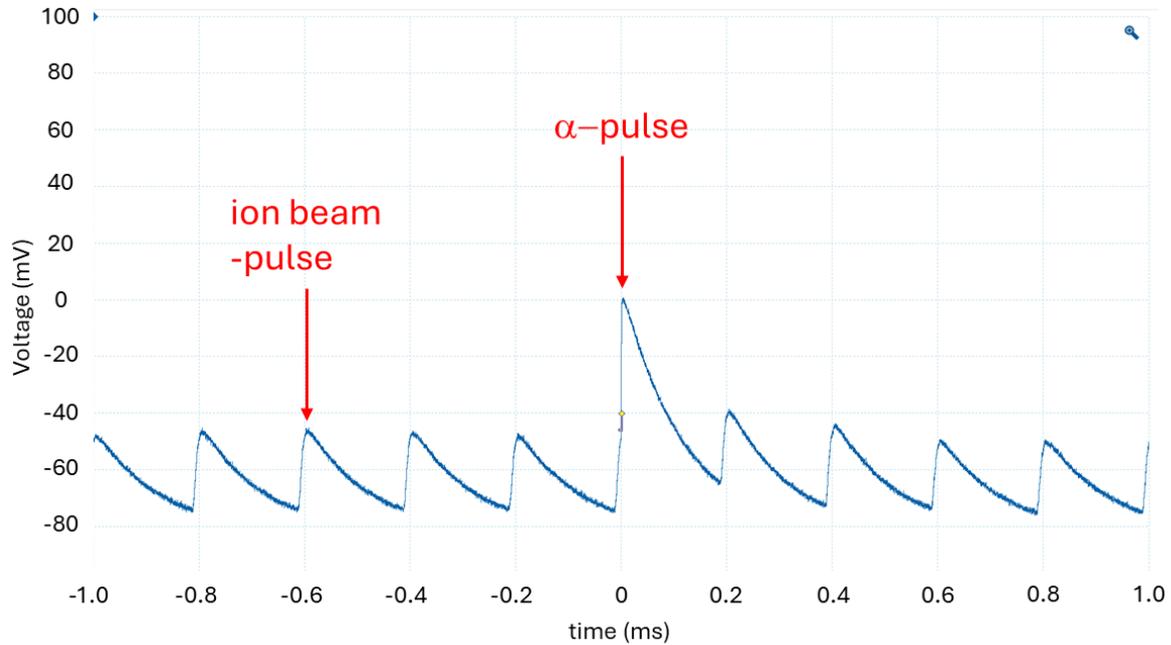

*Figure 3: Waveform of the Si-PIN detector signal from the preamplifier operating in a beam-target experiment with nominal beam energy of 90-keV, 5% duty cycle and 5kHz rep. rate. The alpha-pulse can be seen superimposed on-top of the ion beam pulse noise. With CR-RC filtering, it is possible to deconvolute the alpha pulse from the noise.*

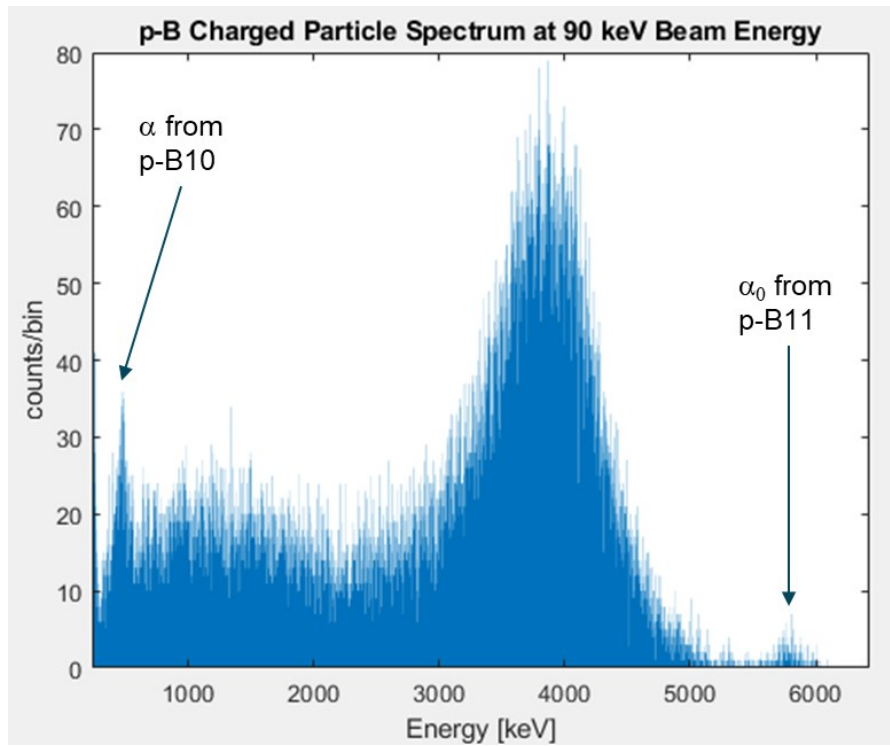

*Figure 4: The spectrum was taken over an 8-hour period. The spectrum matches one that was taken from a more traditional accelerator facility studying the p-B11 reaction [10]. Additionally, due to the lower accelerating potential, the p-B10 mono-energetic alpha can be observed. In addition, we also observe the $a_0$ from the coincident alpha of Be-8.*

**Am-241 Calibration:**

We used Am-241 alpha source to calibrate the linearity of the detector. Using different thickness of foils, between 0 to 20 μm, we're able to attenuate the energy of the alphas to energies from 5.486 MeV down to ~1 MeV. Figure 5 shows the energy spectrum for the different thickness cases. The results were obtained using the CR-RC$^4$ filtering with a time constant of 4 μs described in the previous section. The "no foil" (0 mm) case was used to calibrate the 5.486 MeV Am-241 energy. Table 1 compares the SRIM calculation with the measured peak energy using the detector, showing a strong agreement between the calculated and measured values.

*Table 1: Attenuated Am-241 alpha energy for different foil thicknesses. 2nd column shows the calculated energy using SRIM; 3rd column shows the energy at the right edge of the peak measured with the PIN detector.*

| Foil Thickness [um] | Peak Energy (SRIM) | Peak Energy (edge-cal) |
|---|---|---|
| 0 | **5.486** | **5.486** |
| 4 | 4.814 | 4.858 |
| 8 | 4.151 | 4.165 |
| 12 | 3.373 | 3.394 |
| 16 | 2.479 | 2.571 |
| 20 | 1.397 | 1.315 |

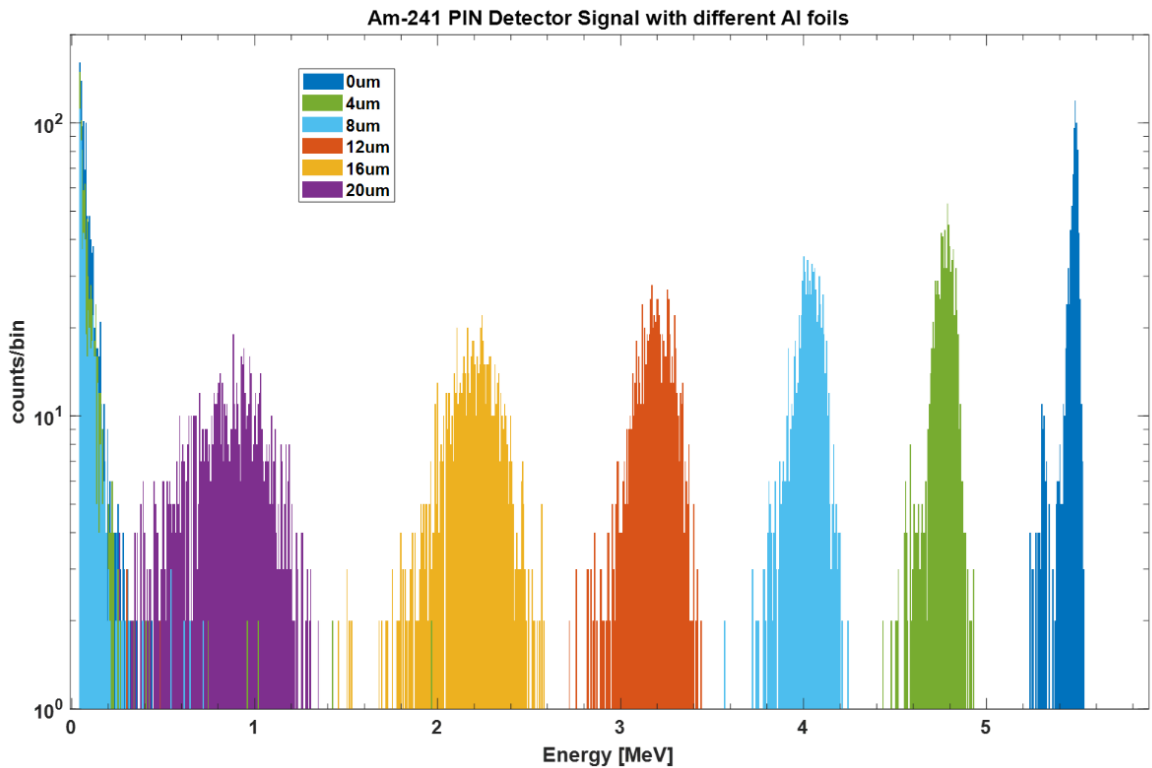

*Figure 5: Am-241 calibration of the Si-PIN detector using various foil thicknesses.*

**Data Acquisition Module**

The data acquisition module was developed by ARI to ensure the entire system is as compact and electrically isolated as possible. A 20 MSPs analog-to-digital converter (Analog Devices AD9629) captures the signal from the PIN detector's preamp output with 12-bit resolution and range of +/-1.8V. The signals are then processed by a high-performance microcontroller unit (STM32U595), operating at a nominal clock speed of 160 MHz with excellent power efficiency. Additionally, operating at +/-1.8 V instead of the typical +/-3.3V further reduces power consumption. Data is transmitted from the data acquisition module to a separate module using optical fiber connections. This separate module converts the optical signal to USB-UART, enabling communication with a computer. There are LEDs on the PCB to indicate both battery status and signal detection. It provides an effective method displaying the status of the unit without needing to operate the computer.

The data transfer workflow is illustrated in Figure 6. The top and bottom sections of the system are electrically isolated, with fiber optics serving as the data link between them. This isolation ensures low-noise operation of the PIN detector and the associated data acquisition electronics. For optical communication, we use standard plastic optical fiber (POF) with a 1000-µm core, 2.2-mm cladding, and connectors sourced from Industrial Fiber Optics Inc. [9]. The system employs an LED (IF-E91D) and a photodetector (IF-D91B) operating in the infrared wavelength range around 870 nm. This configuration was selected for its cost-effectiveness and high baud-rate capability, supporting data transfer speeds of up to 100 Mbps over 10 meters.

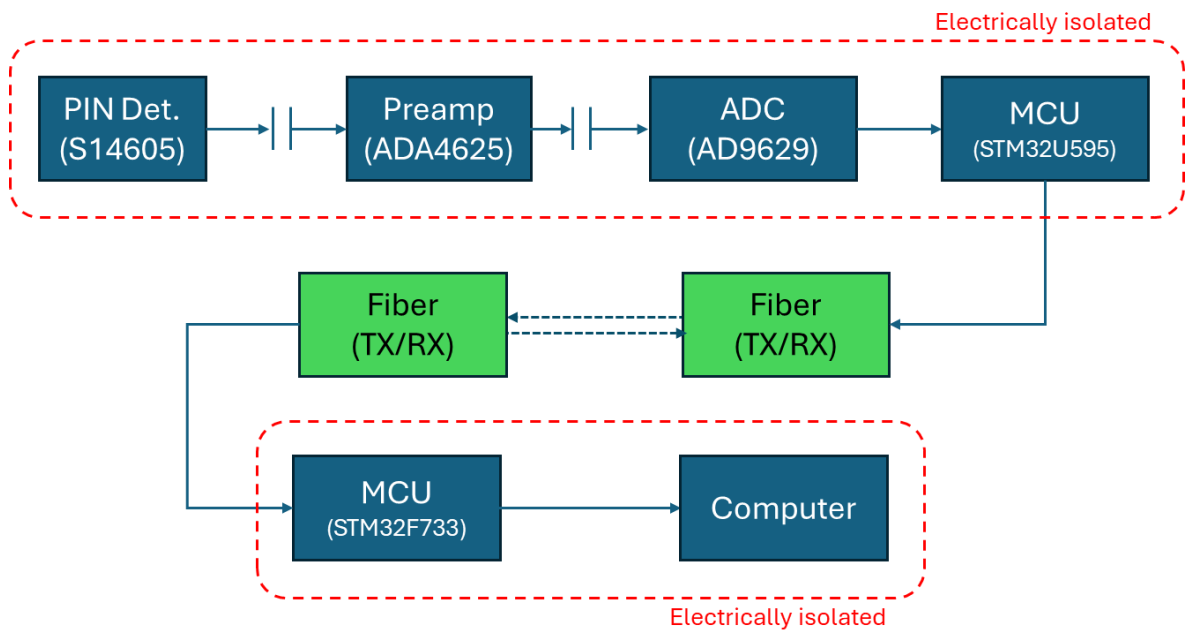

*Figure 6: Signal flow chart of the Si-PIN detector system with optical fiber datalink communication. This allows the entire detection system, including data processing, to be isolated from the receiving computer.*

The integrated PIN detector with a data acquisition module was utilized in the ARI-IBS system to detect charged particles from D-D fusion. The ion beam operates at a nominal voltage of 30 keV, with a beam duty cycle of 5% and a current of 0.05 mA. This output is low enough to ensure user safety from excessive radiation exposure while still producing sufficient fusion events to be reliably detected by the PIN detector. A typical D-D charged particle spectrum is shown in figure 7, where the three distinct charged particles (proton, triton, and helion) from D-D fusion are clearly visible. To generate the spectrum, each pulse from the preamplifier signal was processed using a standard CR-RC$^4$ filter. The third peak in the processed data was normalized to the D-D proton energy of 3.02 MeV. The noise threshold of the spectrum was determined to be approximately 0.1 MeV, which was used as the baseline for calculating the signal-to-noise (S/N) ratios. The calculated S/N ratios for the proton, triton, and helion signals were approximately 30, 10, and 5, respectively. It is worth noting that the helion signal, which has a nominal peak energy of 0.82 MeV, experienced attenuation due to the 0.8 μm aluminum foil used to block optical emissions. This attenuation reduced the detected energy by several hundred keV, resulting in a final measured energy of approximately 560 keV. Similarly, the triton signal is also slightly reduced.

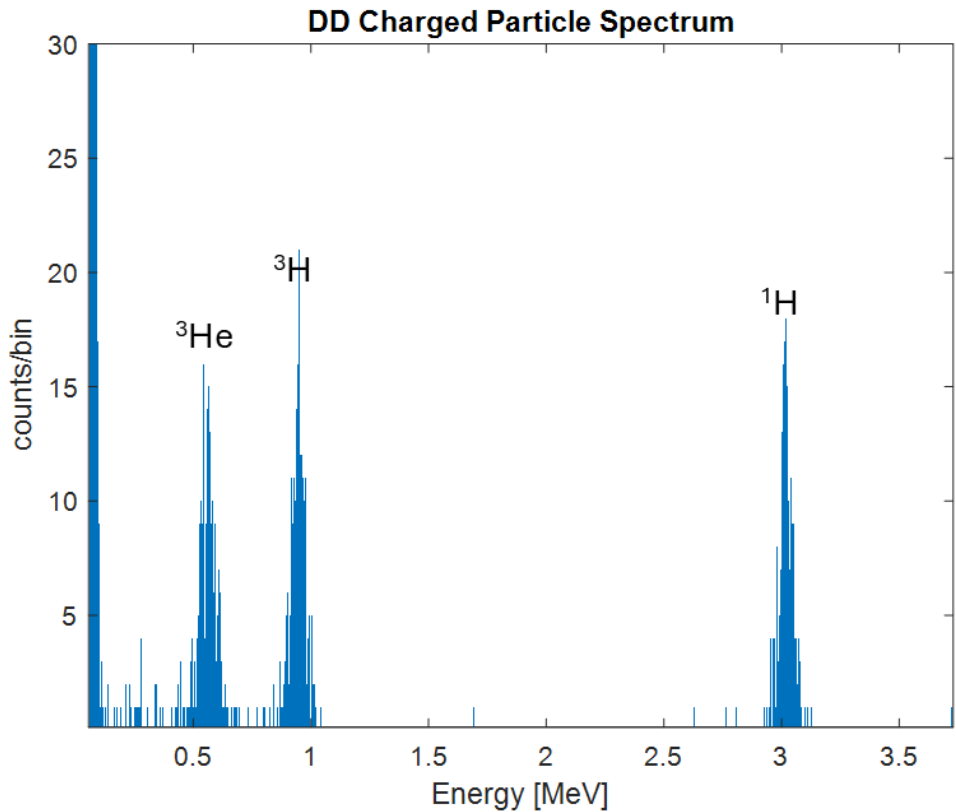

*Figure 7: Typical D-D charged particle spectrum obtained using the Si-PIN detector with integrated data acquisition system. The beam-target system was operated at 30-keV beam energy at 5% duty cycle with nominal current of 0.05mA.*

**Operation in Vacuum**

For future applications, we envision using this device to measure charged particles within large fusion devices. In such environments, the unit must be capable of operating reliably under vacuum conditions, including both the electronics and the batteries. To validate its performance, we tested the entire ARPCB-117 assembly (Fig. 8) in an 8-inch vacuum chamber. The assembly, which includes the PIN detector preamplifier and Li-ion A123 batteries, was connected and positioned such that the PIN detector faced a Po-210 alpha source mounted on one of the chamber ports as a calibration test source. Optical fibers were used to transmit data between the atmospheric and vacuum sides, routed through a CF-2.75 flange. To ensure vacuum compatibility, the TX and RX fibers were securely glued in place with high-vacuum-compatible epoxy. The unit was fully electrically isolated from the chamber, which minimizes conducted noise pickup and enhances signal integrity.

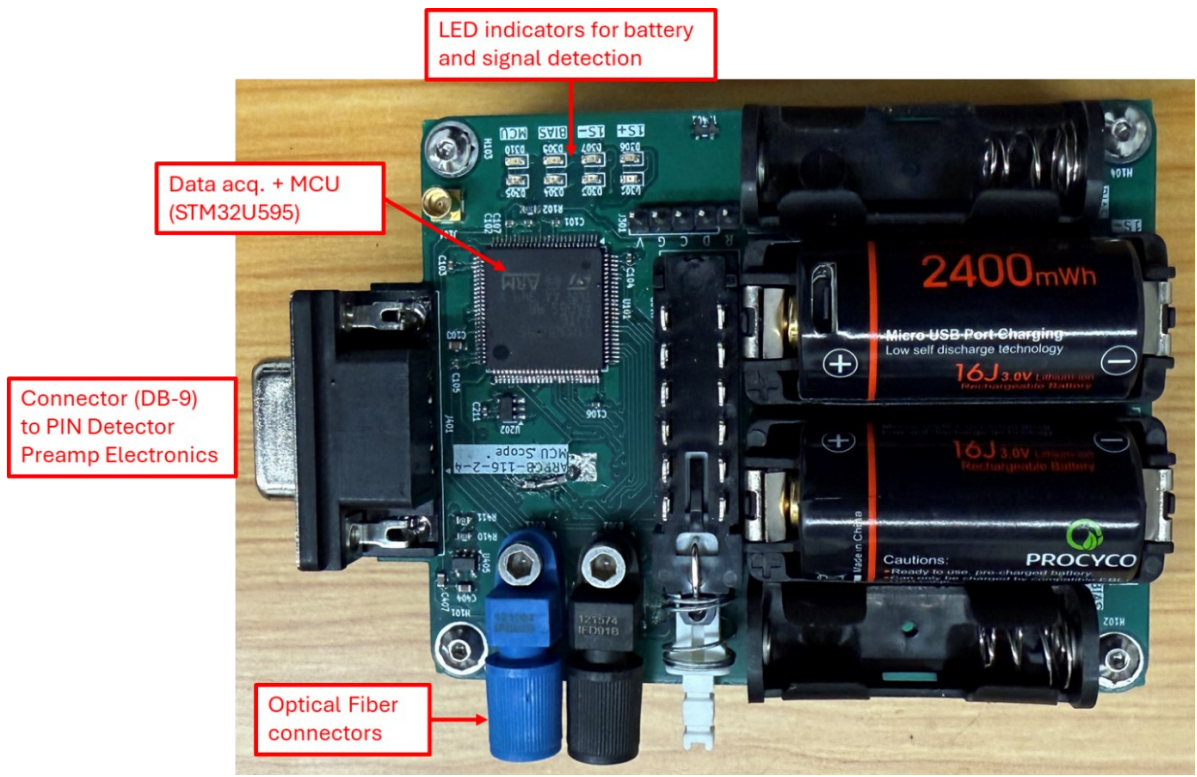

*Figure 8: ARPCB-116 assembly detailing the various components. The unit is electrically isolated with its own battery supply and communicates via. optical fiber.*

We use a standard vacuum pumping system consisting of a turbomolecular pump (Agilent TwisTorr FS 84) backed by a dry diaphragm roughing pump (Agilent MD1) to achieve high vacuum conditions in the chamber (Fig. 9). After approximately 2 hours, the chamber reaches a quasi-ultimate vacuum of ~1.6E-5 Torr. The system was maintained under high vacuum conditions for about 6 hours, during which no abnormalities were observed in the batteries or other electronic components. The vacuum remained stable, consistently below 2E-5 Torr.

Using the PIN detector preamplifier electronics, we successfully observed pulses from the Po-210 alpha source. In high vacuum, the alpha particle signal exhibited a peak height of approximately 70 mV, consistent with expectations as the particles experience no attenuation in the absence of a gas medium. For comparison, under atmospheric conditions, the peak height was reduced to approximately 30 mV due to attenuation from the 2–3 cm of air the particles must travel through (Fig. 10).  Additionally, we tested the system at a gas pressure of 2 Torr and found no significant difference in signal characteristics compared to those observed at ~1E-4 Torr. This demonstrates the robustness of the detector and electronics under varying low-pressure conditions.  It is worth noting that when cycling the unit between atmospheric and vacuum pressure, one should use a bias voltage well below the +150V maximum recommended in order to eliminate the risk of Paschen curve breakdowns.

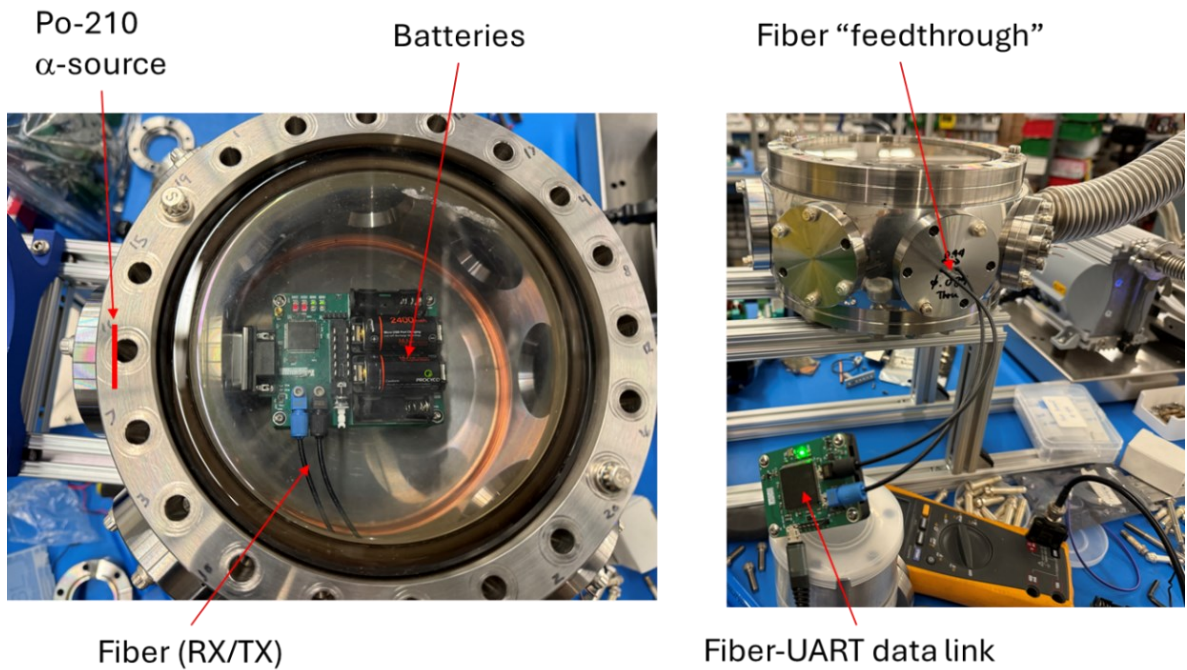

*Figure 9: Testing of the Si-PIN detector and data acquisition system in a high vacuum chamber.  The optical fiber datalink was passed through to atmospheric side to connect with the USB-UART module, which is then connected to the computer.*

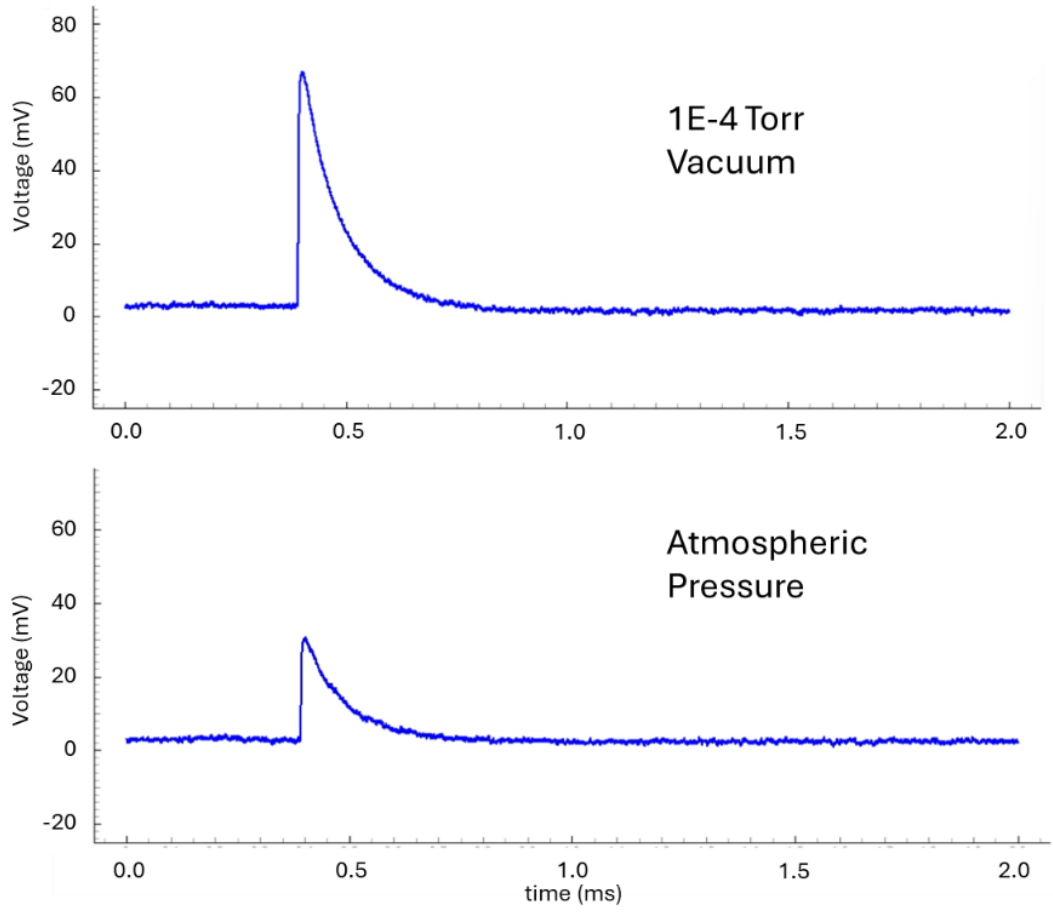

*Figure 10: Si-PIN detector signal from Po-210 under different operating pressures of 1E-4 Torr (top) and atmospheric pressure (bottom). Note the lower peak height observed for the atmospheric pressure signal due to the particle energy being attenuated by the atmospheric gas medium.*

## Conclusion

We have developed a compact PIN detector with integrated electronics, specifically optimized for high-energy particle detection. The system has been successfully used to detect alpha particles from radioactive sources (e.g., Am-241, Po-210) as well as charged particles from D-D fusion reactions. Engineered with a strong emphasis on noise immunity, the detector delivers excellent signal-to-noise ratio performance.  The data link between the detector electronics and the control computer utilizes optical fiber communication, providing complete electrical isolation. This setup enables reliable data transmission over 10 meters without any degradation in the detector's electrical performance.  Furthermore, the entire assembly, including the battery-powered electronics, has been rigorously tested in both high-vacuum conditions (<1E-4 Torr) and low-vacuum environments (2 Torr). Results confirm the system's robustness and reliable operation in vacuum for extended periods, making it well-suited for experimental applications. In future work, we plan to explore and characterize the effect of X-rays on the detector. However, based on our results with the ion beam so far, we have not observed a significant impact on signal integrity.  This compact PIN detector offers a valuable complement to other detection methods, such as solid-state nuclear track detectors (SSNTDs), for accurately verifying the presence and characteristics of fusion products.

## Acknowledgement

The authors are grateful for discussions with and support from Prof. Roger Falcone, Prof. Richard Petrasso, Prof. Jyhpyng Wang, Dr. Hao-Lin Chen, Dr. Kosta Yanev, Dr. Gianluca Gregori, Mr. Peter Liu, Mr. Paul Chau, Mr. James Chen, Dr. Zhe Su, Dr. Nathan Eschbach, Dr. Mason Guffey, Dr. David Chu, Dr. Peter Hsieh, Dr. Steve Hwang, Mr. Wilson Wu, Mr. Charles Wu, Mr. David Noriega, Mr. Ryan Yan, Ms. Lilly Zhang, Ms. Belinda Mei.  This work was conducted by scientists and engineers as employees of Alpha Ring International Limited and its affiliated companies. The study was funded by Alpha Ring International Limited, which supports the education and training of a fusion industry workforce. The funding from Alpha Ring International Limited did not influence the scientific integrity or the results of the study.